\documentstyle[12pt]{article}

\makeatletter
\@addtoreset{equation}{section}
\makeatother

\let\la=\label  
   
 \def\bd{\begin{document}} \def\ed{\end{document}}
\def\ds{\documentstyle} \let\fr=\frac \let\bl=\bigl \let\br=\bigr
\let\Br=\Bigr \let\Bl=\Bigl 
\let\bm=\bibitem
\let\na=\nabla
\let\pa=\partial \let\ov=\overline 
\newcommand{\be}{\begin{equation}} 
\newcommand{\ee}{\end{equation}} 
\def\ba{\begin{array}}
\def\ea{\end{array}}
\newcommand{\ho}[1]{$\, ^{#1}$}
\newcommand{\hoch}[1]{$\, ^{#1}$}
\newcommand{\bea}{\begin{eqnarray}} 
\newcommand{\eea}{\end{eqnarray}} 
\newcommand{\ra}{\rightarrow}
\newcommand{\lra}{\longrightarrow}
\newcommand{\Lra}{\Leftrightarrow}
\newcommand{\ap}{\alpha^\prime}
\newcommand{\bp}{\tilde \beta^\prime}
\newcommand{\tr}{{\rm tr} }
\newcommand{\Tr}{{\rm Tr} } 
\newcommand{\NP}{Nucl. Phys. }
\newcommand{\tamphys}{\it Center for Theoretical Physics\\
Texas A\&M University, College Station, Texas 77843}

\newcommand{\auth}{M. J. Duff}

\begin{document}

\begin{flushright}
\hfill{CTP-TAMU-23/98}\\
\hfill{hep-th/9805177}\\
\end{flushright}

\vspace{24pt}

\begin{center}
{ \large 
{\bf A Layman's Guide to M-Theory\footnote{Talk delivered at the Abdus 
Salam Memorial Meeting, ICTP, Trieste, November 1997. }}.}

\vspace{36pt}

\auth

\vspace{10pt}

{\tamphys}

\vspace{44pt}

\underline{ABSTRACT}

\end{center}
The best candidate for a fundamental unified theory of all physical 
phenomena is no longer ten-dimensional superstring theory but rather
eleven-dimensional {\it M-theory}. In the words of Fields medalist Edward
Witten,  ``M stands for `Magical', `Mystery' or `Membrane', according to
taste''. New evidence in favor of this theory is appearing daily on the
internet and represents the most exciting development in the subject since
1984 when the superstring revolution first burst on the scene.

{\vfill\leftline{}\vfill}

\pagebreak
\setcounter{page}{1}

\section{Abdus Salam}

The death of Abdus Salam was a great loss not only to his
family and to the physics community; it was a loss to all mankind. For he was
not only one of the finest physicists of the twentieth century, having 
unified
two of the four fundamental forces in Nature, but he also dedicated his life 
to
the betterment of science and education in the Third World and to the cause 
of
world peace. Although he won the Nobel Prize for physics, a Nobel Peace Prize 
would have been entirely appropriate.

 At the behest of Patrick Blackett, Salam moved to Imperial College, London, 
 in
1957 where he founded the Theoretical Physics Group.  He remained at Imperial as 
Profesor of Physics for the rest of his carreer. I was fortunate enough
to be his PhD student at Imperial College from 1969 to 1972, and then his 
postdoc at the ICTP from 1972 to 1973.  

Among Salam's earlier achievements was the role played by renormalization 
in quantum field theory when, in particular, he amazed his Cambridge 
contempories
with the resolution of the notoriously thorny problem of overlapping 
divergences.
His brilliance then burst on the scene once more when he proposed the famous
hypothesis that {\it All neutrinos are left-handed}, a hypothesis which
inevitably called for a violation of parity in the weak interactions. He was
always fond of recalling his visit to Switzerland where he
submitted (or should I say ``humbly'' submitted\footnote{``I am a humble man'' was
something of a catchphrase for Salam and used whenever anyone tried to make
physics explanations more complicated than necessary. For more Salam 
anecdotes, including an earlier encounter with Pauli, the reader is 
referred to a forthcoming biographical memoir for the American 
Philosophical Society by Freeman Dyson.}) his two-component 
neutrino idea to the formidable Wolfgang Pauli.
Pauli responded with a note urging Salam to ``think of something 
better''! So Salam delayed publication until
after Lee and Yang had conferred the mantle of respectibility on parity
violation. That taught Salam a valuable lesson and he would constantly 
advise his students never to listen to grand old men\footnote{ I hope this
student, at least, has lived up to that advice!}. It also taught him to adopt a
policy of publish or perish, and his scientific output was prodigious with 
over 300 publications.

Of course, the work that won him the 1979 Nobel Prize that he shared with 
Glashow
and Weinberg was for the electroweak unification which combined several of 
his
abiding interests: renormalizability, non-abelian gauge theories and 
chirality.
His earlier work in 1960 with Goldstone and Weinberg on spontaneous symmetry
breaking and his work with John Ward in the mid 1960s on the weak 
interactions was
no doubt also influential. Looking back on my time as a student 
in
the Theory Group at Imperial from 1969 to 1972, a group that included not 
only
Abdus Salam but also Tom Kibble, one might think that I would have 
been  
uniquely poised to take advantage of the new ideas in spontaneously 
broken gauge theories. Alas, it was not to be since no-one suggested 
that electroweak 
unification would be an interesting topic of research. In fact I did not 
learn about
spontaneous symmetry breaking until after I got my PhD! The reason, of 
course,
is that neither Weinberg nor Salam  (nor anybody else) fully realized the
importance of their model until `t Hooft proved its renormalizability in 1972 and
until the discovery of neutral currents at CERN. Indeed, the Nobel Committee 
was
uncharacteristically prescient in awarding the Prize to Glashow, Weiberg and
Salam in 1979 because the $W$ and $Z$ bosons were not discovered
experimentally at CERN until 1982. Together with Pati, Salam went on to 
propose
that the strong nuclear force might also be included in this unification. 
Among
the predictions of this {\it Grand Unified Theory} are magnetic monopoles and
proton decay: phenomena which are still under intense theoretical and
experimental investigation. More recently, it was Salam, together with his
lifelong collaborator John Strathdee who first proposed the idea of 
superspace, a
space with both commuting and anticommuting coordinates, which underlies all 
of
present day research on supersymmetry.

However, it is to Abdus Salam that I owe a tremendous debt as the man who 
first
kindled my interest in the Quantum Theory of Gravity: a subject which at the 
time
was pursued only by mad dogs and Englishmen. (My thesis title: {\it Problems 
in
the Classical and Quantum Theories of Gravitation} was greeted with hoots of
derision when I announced it at the Cargese Summer School en route to my 
first
postdoc in Trieste. The work originated with a bet between Abdus Salam and
Hermann Bondi about whether you could generate the Schwarzschild solution 
using
Feynman diagrams. You can (and I did) but I never found out if Bondi ever 
paid
up.) It was inevitable that Salam would not rest until the fourth and most
enigmatic force of gravity was unified with the other three.  Such a 
unification
was always Einstein's dream and it remains the most challenging tasks of 
modern
theoretical physics and one which attracts the most able and active 
researchers,
such as those here today.  

I should mention that being a student of someone so bursting with new ideas 
as
Salam was something of a mixed blessing: he would assign a research problem
and then disappear on his travels for weeks at a time\footnote{Consequently, 
it
was to Chris Isham that I would turn for practical help with my PhD thesis.}. On
his return he would ask what you were working on. When you began to explain 
your
meagre progess he would usually say ``No, no, no. That's all old hat. What 
you
should be working on is this'', and he would then allocate a completely new
problem! 

I think it was Hans Bethe who said that there are two kinds of genius. The
first group (to which I would say Steven Weinberg, for example, belongs) 
produce
results of such devastating logic and clarity that they leave you feeling 
that
you could have done that too (if only you were smart enough!). The second 
kind
are the ``magicians'' whose sources of inspiration are completely baffling. 
Salam,
I believe, belonged to this magic circle and there was always an element of
eastern mysticism in his ideas that left you wondering how to fathom his
genius. 

Of course, these scientific achievements reflect only one side of Salam's
character. He also devoted his life to the goal of international peace and
cooperation, especially to the gap between the developed and developing
nations. He firmly believed that this disparity will never be remedied
until the Third World countries become the arbiters of their own scientific 
and
technological destinies. Thus this means going beyond mere financial aid and
the exportation of technology; it means the training of a scientific elite 
who
are capable of discrimination in  all matters scientific. He would thus
vigorously defend the teaching of esoteric subjects such as theoretical
elementary particle physics against critics who complained that the time and
effort would be better spent on agriculture. His establishment of the ICTP in
Trieste was an important first step in this direction.

It is indeed a tragedy that someone so vigorous and full of life as Abdus 
Salam
should have been struck down with such a debilitating disease. He had such a
wonderful {\it joie de vivre} and his laughter, which most resembled a 
barking
sea-lion, would reverberate throughout the corridors of the Imperial College
Theory Group. When the deeds of great men are recalled, one often hears the
clich\'{e} ``He did not suffer fools gladly'', but my memories of Salam at
Imperial College were quite the reverse. People from all over the world would
arrive and knock on his door to expound their latest theories, some of them 
quite
bizarre.  Yet Salam would treat them all with the same courtesy and respect. 
Perhaps it was because his own ideas always bordered on the outlandish that 
he
was so tolerant of eccentricity in others; he could recognize pearls of 
wisdom
where the rest of us saw only irritating grains of sand. A previous example 
was
provided by the military attache from the Israeli embassy in London who
showed up one day with his ideas on particle physics. Salam was impressed
enough to take him under his wing. The man was Yuval Ne'eman and the
result was flavor $SU(3)$.  

Let me recall just one example of a crazy Salam idea.  In that period 1969-72, 
one of the hottest topics was the Veneziano Model and I distinctly remember 
Salam
remarking on the apparent similarity between the mass and angular momentum
relation of a Regge trajectory and that of an extreme black hole.  Nowadays, 
of
course, string theorists will juxtapose black holes and Regge slopes without
batting an eyelid but to suggest that black holes could behave as elementary
particles back in the late 1960's was considered preposterous by minds lesser
than Salam's. (A comparison of the gyromagnetic ratios of spinning black 
holes and
elementary string states is the subject of some of my recent research, so in 
this
 respect Salam was 25 years ahead of his time!)  As an
interesting historical footnote let us recall that at the time Salam had to
change the gravitational constant to match the hadronic scale, an idea which
spawned his {\it strong gravity}; today the fashion is the reverse and we 
change
the Regge slope to match the Planck scale!

Theoretical physicists are, by and large, an honest bunch: occasions when
scientific facts are actually deliberately falsified are almost unheard of.
Nevertheless, we are still human and  consequently want to present our 
results in
the best possible light when writing them up for publication. I recall a 
young
student approaching Abdus Salam for advice on this ethical dilemma: 
``Professor
Salam, these calculations confirm most of the arguments I have been making so
far. Unfortunately, there are also these other calculations which do not 
quite
seem to fit the picture. Should I also draw the reader's attention to these 
at
the risk of spoiling the effect or should I wait? After all, they will 
probably
turn out to be irrelevant.'' In a response which should be immortalized in 
{\it
The Oxford Dictionary of Quotations}, Salam replied: ``When all else fails, 
you
can always tell the truth''.

Amen.  

\section{Magical Mystery Membranes}
\la{Introduction}

Up until 1995, hopes for a final theory 
\cite{Weinberg} that would
reconcile gravity and quantum mechanics, and describe all physical
phenomena, were pinned on {\it superstrings}: one-dimensional objects whose 
vibrational modes represent the elementary particles 
and which live in a ten-dimensional universe \cite{Green}.  All that
has now changed.  In the last two years ten-dimensional superstrings
have been subsumed by a deeper, more profound, new theory:
eleven-dimensional {\it $M$-theory} \cite{Duff}.  The purpose of the present 
paper is to convey to the layman some of this excitement. 

According to the standard model of the strong nuclear, weak nuclear and
electromagnetic forces, all matter is made up of certain building block
particles called {\it fermions} which are held together by force-carrying
particles called {\it bosons}.  This standard model does not incorporate
the gravitational force, however.  A vital ingredient in the quest to go
beyond this standard model and to find a unified theory embracing all
physical phenomena is {\it supersymmetry}, a symmetry which (a) unites the
bosons and fermions, (b) requires the existence of gravity and (c) places
an upper limit of {\it eleven} on the dimension of spacetime.  For these
reasons, in the early 1980s, many physicists looked to eleven-dimensional
{\it supergravity} in the hope that it might provide that elusive
superunified theory \cite{Freedman}.  Then in 1984 superunification
underwent a major paradigm-shift: eleven-dimensional supergravity was
knocked off its pedestal by ten-dimensional superstrings.  Unlike
eleven-dimensional supergravity, superstrings appeared to provide a
quantum consistent theory of gravity which also seemed capable, in
principle, of explaining the standard model\footnote{For an up-to-date 
non-technical 
account of string theory, the reader is referred to the forthcoming popular 
book by Brian Greene \cite{Greene}.}. 

Despite these startling successes, however, nagging doubts persisted about
superstrings. First, many of the most important questions in string
theory, in particular how to confront it with experiment and how to
accommodate quantum black holes, seemed incapable of being answered within
the traditional framework called {\it perturbation theory}, according to
which all quantities of interest are approximated by the first few terms
in a power series expansion in some small parameter. They seemed to call
for some new, {\it non-perturbative}, physics. Secondly, why did there
appear to be {\it five} different mathematically consistent superstring
theories: the $E_8 \times E_8$ heterotic string, the $SO(32)$ heterotic
string, the $SO(32)$ Type $I$ string, the Type $IIA$ and Type $IIB$
strings? If one is looking for a unique {\it Theory of Everything}, this
seems like an embarrassment of riches! Thirdly, if supersymmetry permits
eleven dimensions, why do superstrings stop at ten? This question became
more acute with the discoveries of the {\it supermembrane} in 1987 the {\it
superfivebrane} in 1992. These are bubble-like supersymmetric extended
objects with respectively two and five dimensions moving in an
eleven-dimensional spacetime, which are related to one another by a {\it
duality} reminiscent of the electric/magnetic duality that relates an
electric monopole ( a particle carrying electric charge) to a magnetic
monopole (a hypothetical particle carrying magnetic charge).  Finally,
therefore, if we are going to generalize zero-dimensional point particles
to one-dimensional strings, why stop there? Why not two-dimensional
membranes or more generally $p$-dimensional objects (inevitably dubbed
{\it $p$-branes})? In the last decade, this latter possibility
of spacetime bubbles was actively pursued by a small but dedicated group of
theorists \cite{Duffsutton}, largely ignored by the orthodox superstring
community.

\begin{table}
\[
\left.
\begin{array}{ll}
E_{8}¥\times E_{8}¥~~ heterotic~~ string &\\
SO(32)~~ heterotic~~ string &\\
SO(32)~~ Type~~I~~string & \\
Type~~ IIA ~~string&\\
Type~~ $IIB$ ~~string&
\end{array}
\right \}M~~theory
\]
\la{M}
\caption{The five apparently different string theories are really 
just different corners of M-theory.}
\end{table}

Although it is still too early to claim that all the problems of
string theory have now been resolved, $M$-theory seems a big step in the
right direction. First, it is intrinsically non-perturbative and
already suggests new avenues both for particle physics and black hole
physics.  Secondly, it is is an eleven-dimensional theory which, at
sufficently low energies, looks, ironically enough, like
eleven-dimensional supergravity. Thirdly, it subsumes all five consistent
string theories and shows that the
distinction we used to draw between them is just an artifact of
perturbation theory. See Table 1.  Finally, it incorporates 
supermembranes  and that is
why $M$ stands for Membrane. However, it may well be that we are only just
beginning to scratch the surface of the ultimate meaning of $M$-theory,
and for the time being therefore, $M$ stands for Magic and Mystery too. 

\section{Symmetry and supersymmetry}

Central to the understanding of modern theories of the fundamental forces
is the idea of {\it symmetry}: under certain changes in the way we
describe the basic quantities, the laws of physics are nevertheless seen
to remain unchanged. For example, the result of an experiment should be
the same whether we perform it today or tomorrow; this symmetry is called
{\it time translation invariance}. It should also be the same before and
after rotating our experimental apparatus; this symmetry is called {\it
rotational invariance}. Both of these are examples of {\it spacetime
symmetries}. Indeed, Einstein's general theory of relativity is based on
the requirement that the laws of physics should be invariant under {\it
any} change in the way we describe the positions of events in spacetime. 
In the standard model of the strong, weak and electromagnetic forces there
are other kinds of {\it internal} symmetries that allow us to change the
roles played by different elementary particles such as electrons and
neutrinos, for example. These statements are made precise using the branch
of mathematics known as {\it Group Theory}.  The standard model is based
on the group $SU(3) \times SU(2) \times U(1)$, where $U(n)$ refers to {\it
unitary} $n\times n$ matrices and $S$ means unit determinant.  Grand
Unified Theories, which have not yet received the same empirical support
as the standard model, are even more ambitious and use bigger groups, such
as $SU(5)$, which contain $SU(3)\times SU(2) \times U(1)$ as a subgroup. 
In this case, the laws remain unchanged even when we exchange the roles of
the {\it quarks} and electrons. Thus it is that the greater the
unification, the greater the symmetry required. The standard model
symmetry replaces the three fundamental forces: strong, weak and
electromagnetic, with just two: the strong and electroweak. Grand unified
symmetries replace these two with just one strong-electroweak force. In
fact, it is not much of an exaggeration to say that the search for the
ultimate unified theory is really a search for the right symmetry.  

At this stage, however, one might protest that some of these internal
symmetries fly in the face of experience. After all, the electron is very
different from a neutrino: the electron has a non-zero mass whereas the
neutrino is massless\footnote{Or at most has a very tiny mass.}.  Similarly, 
the electrons which orbit the  atomic
nucleus are very different from the quarks out of which the protons and
neutrons of the nucleus are built. Quarks feel the strong nuclear force
which holds the nucleus together, whereas electrons do not. These feelings
are, in a certain sense, justified: the world we live in does not exhibit
the $SU(2) \times U(1)$ of the standard model nor the $SU(5)$ of the grand
unified theory. They are what physicists call ``broken symmetries''.  
The idea is that these theories may exist in several different {\it
phases}, just as water can exist in solid, liquid and gaseous phases. In
some of these phases the symmetries are broken but in other phases, they
are exact. The world we inhabit today happens to correspond to the
broken-symmetric phase, but in conditions of extremely high energies or
extremely high temperatures, these symmetries may be restored to their
pristine form.  The early stages of our universe, shortly after the Big
Bang, provide just such an environment. Looking back further into the
history of the universe, therefore, is also a search for greater and
greater symmetry. The ultimate symmetry we are looking for may well be the
symmetry with which the Universe began.

$M$-theory, like string theory before it, relies crucially on the idea,
first put forward in the early 1970s, of a spacetime {\it supersymmetry}
which exchanges bosons and fermions.  Just as the earth rotates on its own
axis as it orbits the sun, so electrons carry an intrinsic angular
momentum called ``spin'' as they orbit the nucleus in an atom.  Indeed all
elementary particles carry a spin $s$ which obeys a quantization rule
$s=nh/4\pi$ where $n=0,1,2,3,...$ and $h$ is Planck's constant. Thus
particles may be divided into {\it bosons} or {\it fermions} according as
the spin, measured in units of $h/2\pi$, is integer $0,1,2,...$ or
half-odd-integer $1/2,3/2,5/2..$.  Fermions obey the {\it Pauli exclusion
principle}, which states that no two fermions can occupy the same quantum
state, whereas bosons do not. They are said to obey opposite {\it
statistics}. According to the standard model, the quarks and leptons which
are the building blocks of all matter, are spin $1/2$ fermions; the
gluons, W and Z particles and photons which are the mediators of the
strong, weak and electromagnetic forces, are spin $1$ bosons and the Higgs
particle which is responsible for the breaking of symmetries and for
giving masses to the other particles, is a spin $0$ boson. Unbroken
supersymmetry would require that every elementary particle we know of
would have an unknown super-partner with the same mass but obeying the
opposite statistics: for each boson there is a fermion; for each fermion a
boson.  Spin $1/2$ quarks partner spin $0$ {\it squarks}, spin $1$ photons
partner spin $1/2$ {\it photinos}, and so on. In the world we inhabit, of
course, there are no such equal mass partners and bosons and fermions seem
very different. Supersymmetry, if it exists at all, is clearly a broken
symmetry and the new supersymmetric particles are so heavy that they have
so far escaped detection. At sufficiently high energies, however,
supersymmetry may be restored. Supersymmetry may also solve the so-called
{\it gauge hierarchy problem}: the energy scale at which the grand unified
symmetries are broken is vastly higher from those at which the electroweak
symmetries are broken. This raises the puzzle of why the electrons,
quarks, and W-bosons have their relatively small masses and the extra
particles required by grand unification have their enormous masses. Why do
they not all slide to some common scale? In the absence of supersymmetry,
there is no satisfactory answer to this question, but in a supersymmetric
world this is all perfectly natural. The greatest challege currently
facing high-energy experimentalists at Fermi National Laboratory
(Fermilab) in Chicago and the European Centre for Particle Physics (CERN)
in Geneva is the search for these new supersymmetric particles. The
discovery of supersymmetry would be one of the greatest experimental
achievements of the century and would completely revolutionize the way we
view the physical world\footnote{This will present an interesting dilemma 
for 
those pundits who are predicting the {\it End of Science} on
the grounds that all the important discoveries have already been 
made.  Presumably, they will say ``I told you so'' if supersymmetry is not 
discovered, and ``See, there's
one thing less left to discover'' if it is.}.

Symmetries are said to be {\it global} if the changes are the same
throughout spacetime, and {\it local} if they differ from one point to
another. The consequences of {\it local} supersymmetry are even more
far-reaching: it predicts gravity. Thus if Einstein had not already 
discovered General Relativity, local supersymmetry would have forced 
us to invent it. In fact, we are forced to a {\it supergravity} in which the 
graviton, a spin $2$ boson
that mediates the gravitational interactions, is partnered with a spin
$3/2$ {\it gravitino}. This is a theorist's dream because it confronts the
problem from which both general relativity and grand unified theories shy
away: neither takes the other's symmetries into account. Consequently,
neither is able to achieve the ultimate unification and roll all four
forces into one. But local supersymmetry offers just such a possiblity,
and it is this feature above all others which has fuelled the theorist's
belief in supersymmetry in spite of twenty-five years without experimental
support. 

\section{Eleven-dimensional supergravity}

Supergravity has an even more bizarre feature, however, it places an upper
limit on the dimension of spacetime! We are used to the idea that space
has three dimensions: height, length and breadth; with time providing the
fourth dimension of {\it spacetime}. Indeed this is the picture that
Einstein had in mind in 1916 when he proposed general relativity.  But in 
the early 1920's, in their attempts to unify Einstein's gravity and
Maxwell's electromagnetism, Theodore Kaluza and Oskar Klein suggested that
spacetime may have a hidden fifth dimesion.  This idea was quite
succesful: Einstein's equations in five dimensions not only yield the
right equations for gravity in four dimensions but Maxwell's equations
come for free. Conservation of electric charge is just conservation of
momentum in the fifth direction. In order to explain why this extra
dimension is not apparent in our everyday lives, however, it would have to
have a different {\it topology} from the other four and be very small. 
Whereas the usual four coordinates stretch from minus infinity to plus
infinity, the fifth cordinate would lie between $0$ and $2\pi R$. In other
words, it describes a circle of radius $R$. To get the right value for the
charge on the electron, moreover, the circle would have to be tiny, 
$R\sim 10^{-35}~{\rm meters}$, which satisfactorily explains why we are
unaware of its existence. It is difficult to envisage a spacetime with such
a topology but a nice analogy is provide by a garden hose: at large
distances it looks like a line  but closer inspection reveals that at each
point of the line, there is a little circle. So it was that Kaluza and
Klein suggested there is a little circle at each point of four-dimensional
spacetime.  Moreover, this explained for the first time the empirical fact
that all particles come with an electric charge which is an integer
multiple of the charge on the electron, in other words, why electric
charge is {\it quantized}.   

The Kaluza-Klein idea was forgotten for many years but was revived in the
early 1980s when it was realized by Eugene Cremmer, Bernard Julia and
Joel Scherk from the Ecole Normale in Paris that supergravity not only
permits up to seven extra dimensions, but in fact takes its simplest and
most elegant form when written in its full eleven-dimensional glory.
Moreover, the kind of four-dimensional picture we end up with depends on
how we {\it compactify} these extra dimensions: maybe seven of them would
allow us to derive, a la Kaluza-Klein, the strong and weak forces as well
as the electromagnetic.  In the end, however, eleven dimensional
supergravity fell out of favor for several reasons.  

First, despite its extra dimensions and despite its supersymmetry,
eleven-dimensional supergravity is still a {\it quantum field
theory} and runs into the problem from which all such theories suffer: the
quantum mechanical probability for certain processes yields the answer {\it
infinity}. By itself, this is not necessarily a disaster. This problem was
resolved in the late 1940s in the context of Quantum Electrodynamics
(QED), the study of the electromagnetic interactions of photons and
electrons, by showing that these infinities could be absorbed in to a
redefinition or {\it renormalization} of the parameters in the theory such
as the mass and charge of the electron. This {\it renormalization}
resulted in predictions for physical observables which were not only
finite but in spectacular agreement with experiment. Spurred on by the
success of QED, physicists looked for renormalizable quantum field
theories of the weak and strong nuclear interactions which in the 1970s
culminated in the enormously successful standard model that we know
today.  One might be tempted, therefore, to conclude that {\it
renormalizability}, namely the ability to absorb all infinities into a
redefinition of the parameters in the theory, is a prerequisite for any
sensible quantum field theory. However, the central quandary of all
attempts to marry quantum theory and gravity, such as eleven-dimensional
supergravity,  is that Einstein's general theory of relativity
turns out {\it non-renormalizable}! Does this mean that Einstein's theory
should be thrown on the scrapheap?  Actually, the modern view of
renormalizability is a little more forgiving. Suppose we have a
renormalizable quantum field theory describing both light particles and
heavy particles of mass $m$. Even such a renormalizable theory can be made
to look non-renormalizable if we eliminate the heavy particles by using
their equations of motion. The resulting equations for the light particles
are then non-renormalizable but perfectly adequate for describing
processes at energies less than $mc^2$, where $c$ is the velocity of
light. We  run into trouble only if we try to extrapolate them beyond this
range of validity, at which point we should instead resort to the original
version of the theory with the massive particles put back in. In this
light, therefore, the modern view of Einstein's theory is that it is
perfectly adequate to explain gravitational phenomena at low energies but
that at high energies it must be replaced by some more fundamental theory
containing massive particles. But what is this energy, what are these
massive particles and what is this more fundamental theory?

There is a natural energy scale associated with any quantum theory of
gravity. Such a theory combines three ingredients each with their own
fundamental constants: Planck's constant $h$ (quantum mechanics), the
velocity of light $c$ (special relativity) and Newton's gravitational
constant $G$ (gravity). From these we can form the so-called Planck mass 
$m_P=\sqrt{hc/G}$, equal to about $10^{-8}$ kilograms, and the Planck
energy $m_Pc^2$, equal to about $10^{19}$ GeV. (GeV is short for
giga-electron-volts=$10^9$ electron-volts, and an electron-volt is the
energy required to accelerate an electron through a potential difference
of one volt. )  From this we conclude that the energy at which Einstein's
theory, and hence eleven-dimensional supergravity, breaks down is the
Planck energy. On the scale of elementary particle physics, this energy is
enormous\footnote{For this reason,
incidentally, the {\it End of Science} brigade like to claim that, even if
we find the right theory of quantum gravity, we will never be able to test
it experimentally! As I will argue shortly, however, this view is
erroneous.}: the world's most powerful particle acclerators can currently
reach energies of only $10^{4}$ GeV.  So it seemed in the early 1980s that 
we 
were looking for a
fundamental theory which reduces to Einstein's gravity at low energies,
which describes Planck mass particles and which is supersymmetric.
Whatever it is, it cannot be a quantum field theory because we already
know all the supersymmetric ones and they do not fit the bill. 

Equally puzzling was that an important feature of the real world
which is incorporated into both the standard model and grand unified
theories is that Nature is {\it chiral}: the weak nuclear force distinguishes 
between right and left. (As Salam had noted with his left-handed 
neutrino hypothesis). However, as emphasized by Witten
among others, it is impossible via conventional Kaluza-Klein techniques to
generate a chiral theory from a non-chiral one and unfortunately,
eleven-dimensional supergravity, in common with any {\it odd}-dimensional
theory, is itself non-chiral.

\section{Ten-dimensional superstrings}

For both these reasons, attention turned to ten-dimensional superstring
theory. The idea that the fundamental stuff of the universe might
not be pointlike elementary particles, but rather one-dimensional
strings had been around from the early 1970s. Just like violin strings,
these relativistic strings can vibrate and each elementary particle:
graviton, gluon, quark and so on, is identified with a different mode of
vibration. However, this means that there are {\it infinitely many}
elementary particles. Fortunately, this does not contradict experiment because
most of them, corresponding to the higher modes of vibration, will have
masses of the order of the Planck mass and above and will be unobservable
in the direct sense that we observe the lighter ones. Indeed, an infinite
tower of Planck mass states is just what the doctor ordered for curing the
non-renormalizability disease. In fact, because strings are {\it
extended}, rather than pointlike, objects, the quantum mechanical
probabilities involved in string processes are actually {\it finite}. 
Moreover,  when we take the {\it low-energy limit} by eliminating these
massive particles through their equations of motion, we recover a
ten-dimensional  version of supergravity which incorporates Einstein's
gravity. Now ten-dimensional quantum field theories, as opposed to
eleven-dimensional ones, also admit the possibility of {\it chirality}.
The reason that everyone had still not abandoned eleven-dimensional
supergravity in favor of string theory, however, was that the
realistic-looking Type $I$ string, which incorporated internal symmetry
groups containing the $SU(3) \times SU(2) \times U(1)$ of the standard
model, seemed to suffer from inconsistencies or {\it anomalies}, whereas
the consistent non-chiral Type $IIA$ and chiral Type $IIB$ strings did not
seem realistic.     

Then came the September 1984 superstring revolution. First, Michael Green 
from QueenMary and Westfield College, London, and John Schwarz
from the California Institute of Technology showed that the Type $I$ string 
was
free of anomalies provide the group was uniquely $SO(32)$ where O(n) stands
for {\it orthogonal} $n \times n$ matrices. They suggested that a string
theory based on the exceptional group $E_8 \times E_8$ would also have
this property. Next, David Gross, Jeffrey Harvey, Emil Martinec and
Ryan Rohm from Princeton University discovered a new kind of heterotic
(hybrid) string theory based on just these two groups: the 
$E_8 \times E_8$ heterotic string and the $SO(32)$ heterotic string, thus
bringing to {\it five} the number of consistent string theories. Thirdly,
Philip Candelas from the University of Texas, Austin, Gary Horowitz and
Andrew Strominger from the University of California, Santa Barbara
and Witten showed that these heterotic string theories admitted a
Kaluza-Klein compactification from ten dimensions down to four. The
six-dimensional compact spaces belonged to a class of spaces known to the
pure mathematicians as {\it Calabi-Yau manifolds}. The resulting
four-dimensional theories resembled quasi-realistic grand unfied theories
with chiral representations for the quarks and leptons! Everyone dropped 
eleven-dimensional supergravity like a hot brick.  The mood of the times
was encapsulated by Nobel Laureate Murray Gell-Mann in his closing
address at the 1984 Santa Fe Meeting, when he said: ``Eleven Dimensional
Supergravity (Ugh!)''. 

\section{Ten to eleven: it is not too late}

After the initial euphoria, however, nagging doubts about string theory
began to creep in.

Theorists love {\it uniqueness}; they like to think that the
ultimate {\it Theory of Everything} \cite{Weinberg} will one day be
singled out, not merely because all rival theories are in disagreement
with experiment, but because they are mathematically inconsistent. In
other words, that the universe is the way it is because it is the only
possible universe. But string theories are far from unique. Already in ten
dimensions there are five mathematically consistent theories: the Type $I$
$SO(32)$, the heterotic $SO(32)$, the heterotic $E_8\times E_8$, the Type
$IIA$ and the Type $IIB$. (Type $I$ is an {\it open} string in that its
ends are allowed to move freely in spacetime; the remaining four are {\it
closed strings} which form a closed loop.) Thus the first problem is
the {\it uniqueness problem}.  

The situation becomes even worse when we consider compactifying
the extra six dimensions. There seem to be billions of different ways of
compactifying the string from ten dimensions to four (billions of
different Calabi-Yau manifolds) and hence billions of competing
predictions of the real world (which is like having no predictions at
all). This aspect of the uniqueness problem is called the {\it
vacuum-degeneracy problem}. One can associate with each different
phase of a physical system a {\it vacuum state}, so called because it
is the quantum state corresponding to no real elementary particles at
all.  However, according to quantum field theory, this vacuum is
actually buzzing with virtual particle-antiparticle pairs that are
continually being created and destroyed and consequently such vacuum
states carry energy. The more energetic vacua, however, should be unstable
and eventually decay into a (possibly unique) stable vacuum with the least
energy, and this should describe the world in which we live.
Unfortunately, all these Calabi-Yau vacua have the same energy and the
string seems to have no way of preferring one to the other.  By focussing
on the fact that strings are formulated in ten spacetime dimensions and
that they unify the forces at the Planck scale,  many critics of string
theory fail to grasp this essential point. The problem is not so much that
strings are unable to produce four-dimensional models like the standard
model with quarks and leptons held together by gluons, $W$-bosons, $Z$
bosons and photons and of the kind that can be tested experimentally in
current or forseeable accelerators. On the contrary, string theorists can
dream up literally billions of them! The problem is that they have no way
of discriminating between them. What is lacking is some dynamical mechanism
that would explain why the theory singles out one particular Calabi-Yau
manifold and hence why we live in one particular vacuum; in other words,
why the world is the way it is.  Either this problem will not be solved,
in which case string theory will fall by the wayside like a hundred other
failed theories, or else it will be solved and string theory will be put
to the test experimentally. Neither string theory nor $M$-theory is
relying for its credibility on building thousand-light-year accelerators
capable of reaching the Planck energy, as some {\it End-of-Science}
Jeremiahs have suggested.

Part and parcel of the vacuum degeneracy problem is the {\it
supersymmetry-breaking problem}. If superstrings are to describe our world
then supersymmetry must be broken, but the way in which strings achieve
this, and at what energy scales, is still a great mystery.

A third aspect of vacuum degeneracy is the {\it cosmological constant}
problem. Shortly after writing down the equations of general relativity,
Einstein realized that nothing prevented him from adding an extra term,
called the {\it cosmological term} because it affects the rate at which
the universe as a whole is expanding. Current astrophysical data
indicates that the coefficient of this term, called the {\it cosmological
constant}, is zero or at least very small.  Whenever an {\it a priori}
allowed term in an equation seems to be absent, however, theorists always
want to know the reason why. At first sight supersymmetry seems to
provide the answer. The cosmological constant measures the energy of the
vacuum, and in supersymmetric vacua the energy coming from virtual bosons
is exactly cancelled by the energy coming from virtual fermions! 
Unfortunately, as we have already seen, the vacuum in which our universe
currently finds itself can at best have broken supersymmetry and so all
bets are off. As with cake, we can't have our cosmological constant and eat
it too! In common with all other theories one can think of, superstrings as
yet provide no resolution of this paradox. 

On the subject of gravity, let us not forget {\it black holes}. 
According to Cambridge University's Stephen Hawking, they are not as black
as they are painted: quantum black holes radiate energy and hence grow
smaller. Moreover, they radiate energy in the same way irrespective of
what kind of matter went to make up the black hole in the first place. The
rate of radiation increases with diminishing size and the black hole
eventualy explodes leaving nothing behind, not even the grin on the
Cheshire Cat. All the information about the original constituents of the
black hole has been lost and this leads to the {\it information loss
paradox} because such a scenario flies in the face of traditional quantum
mechanics.  On a more pragmatic level, another unsolved problem was that
the thermodynamic entropy formula of the black hole radiation, first
written down by Jacob Bekenstein (Hebrew University), had never received a
{\it microscopic} explanation.  The entropy of a system is a measure of its
disorder, and is related to the number of quantum states that the system
is allowed to occupy.  For a black hole, this number seems incredibly high
but what microscopic forces are at work to explain this? Not even strings,
with their infinite number of vibrational modes seemed to have this
capability.

Given all the good news about string theory, though, string enthusiasts
were reluctant to abandon the theory notwithstanding all these problems.
Might the faults lie not with the theory itself but rather with the
way the calculations are carried out? In common with the standard model
and grand unified theories, the equations of string theory are just too
complicated to solve exactly. We have to resort to an approximation scheme
and the time-honored way of doing this in physics is {\it perturbation
theory}.  Let us recall quantum electrodynamics, for example, and denote
by $e$ the electric charge on an electron.  The ratio $\alpha=2{\pi}
e^2/hc$ is a dimensionless number called, for historical reasons, the fine
structure constant. Fortunately for physicists, $\alpha$ is about $1/137$:
much less than $1$. Consequently, if we can express processes (such as the
probability of one electron scattering off another) in a power series in
this coupling constant $\alpha$, then we can be confident that keeping just
the first few terms in the series will be a good approximation to the exact
result.  As a simple example of approximating a mathematical function
$f(x)$ by a power series, consider 
\be
f(x)=(1-x)^{-1}=1+x+x^2+x^3...
\ee
Provided $x$ is very much less than unity, the first few terms provide
a good approximation.  This is precisely what Richard Feynman was doing
when he devised his {\it Feynman diagram} technique. The same perturbative
techniques work well in the weak interactions where the corresponding
dimensionless coupling constant is about $10^{-5}$. Indeed, this is how the
weak interactions justify their name. When we come to the strong
interactions, however, we are not so lucky. Now the strong fine structure
constant which governs the strength of low-energy nuclear processes, for
example, is of order unity and perturbation theory can no longer be
trusted: each term in the power series expansion is just as big as the
others.  The whole industry of {\it lattice gauge theory} is devoted to an
attempt to avoid perturbation theory in the strong interactions by doing
numerical simulations on supercomputers. It has proved enormously
difficult.  

The point to bear in mind, however, (and one that even string theorists
sometimes forget) is that ``God does not do perturbation theory''; it is
merely a technique dreamed up by poor physicists because it is the best
they can do. Furthermore, although theories such as quantum electrodynamics
manage to avoid it, there is a possible fatal flaw with perturbation
theory.  What happens if the process we are interested in depends
on the  coupling constant in an intrinsically non-perturbative way
which does not even admit a power series expansion? Such
mathematical functions are not difficult to come by: the function 
\be
f(x)=e^{-1/x^2},       
\ee
for example, cannot be approximated by a power series in $x$ no matter
how small $x$ happens to be.  The equations of string theory
are sufficiently complicated that such non-perturbative behaviour cannot be
ruled out.  If so, might our failure to answer the really difficult
problems be more the fault of string {\it theorists} than string
{\it theory}? 

An apparently different reason for having mixed feelings about
superstrings, of course, especially for those who had been pursuing
Kaluza-Klein supergravity prior to the 1984 superstring revolution, was the
dimensionality of spacetime. If supersymmetry permits eleven spacetime
dimensions, why should the theory of everything stop at ten? 

This problem rose to the surface again in 1987 when Eric Bergshoeff of
the University of Groningen, Ergin Sezgin, now at Texas A\&M
University, and Paul Townsend from the University of Cambridge  discovered
{\it The eleven-dimensional supermembrane}.  This membrane is a bubble-like
extended object with two spatial dimensions which moves in a spacetime
dictated by our old friend: eleven-dimensional supergravity!  Moreover,
Paul Howe (King's College, London University), Takeo Inami (Kyoto
University), Kellogg Stelle (Imperial College) and I were then able to
show that if one of the eleven dimensions is a circle, then we can wrap
one of the membrane dimensions around it so that, if the radius of the
circle is sufficiently small, it looks like a string in ten dimensions. In
fact, it yields precisely the Type $IIA$ superstring. This suggested
to us that maybe the eleven-dimensional theory was the more fundamental
after all.  

\section{Supermembranes}

Membrane theory has a strange history which goes back even further than
strings. The idea that the elementary particles might correspond to
modes of a vibrating membrane was put forward originally in 1960 by the
British Nobel Prize winning physicist Paul Dirac, a giant of twentieth
century science who was also responsible for two other daring postulates:
the existence of {\it anti-matter} and the existence of {\it magnetic
monopoles}. Anti-particles carry the same mass but opposite charge from
particles and were discovered experimentally in the 1930s. Magnetic
monopoles carry a single magnetic charge and to this day have not yet been
observed. As we shall see, however, they do feature prominently in
$M$-theory. When string theory came along in the 1970s, there were some
attempts to revive Dirac's membrane idea but without much success. The
breakthrough did not come until 1986 when James Hughes, James Liu and
Joseph Polchinski of the University of Texas showed that, contrary to the
expectations of certain string theorists, it was possible to combine
the membrane idea with supersymmetry: the {\it supermembrane} was born.

Consequently, while all the progress in superstring theory was being made a
small but enthusiastic group of theorists were posing a seemingly very
different question: Once you have given up $0$-dimensional particles in
favor of $1$-dimensional strings, why not $2$-dimensional membranes or in
general $p$-dimensional objects (inevitably dubbed {\it $p$-branes})? 
Just as a $0$-dimensional particle sweeps out a $1$-dimensional {\it
worldline} as it evolves in time, so a $1$-dimensional string sweeps out a
$2$-dimensional {\it worldsheet} and a $p$-brane sweeps out a
$d$-dimensional {\it worldvolume}, where $d=p+1$.  Of course, there must
be enough room for the $p$-brane to move about in spacetime, so $d$ must
be less than the number of spacetime dimensions $D$.  In fact supersymmetry
places further severe restrictions both on the dimension of the extended
object and the dimension of spacetime in which it lives. One can represent
these as points on a graph where we plot spacetime dimension $D$
vertically and the $p$-brane dimension $d=p+1$ horizontally. This graph is
called the {\it brane-scan}. See Table \ref{branescan}. Curiously enough, 
the maximum
spacetime dimension permitted is eleven, where Bergshoeff, Sezgin and
Townsend found their $2$-brane. In the early 80s Green and  Schwarz had
showed that spacetime supersymmetry allows classical superstrings moving
in spacetime dimensions $3,4,6$ and $10$. (Quantum considerations rule out
all but the ten-dimensional case as being truly fundamental. Of course
some of these ten dimensions could be curled up to a very tiny size in the
way suggested by Kaluza and Klein. Ideally six would be compactified in
this way so as to yield the four spacetime dimensions with which we are
familiar.) It was now realized, however, that there were twelve points on
the scan which fall into four sequences ending with the superstrings or
$1$-branes in $D=3,4,6$ and $10$, which were now viewed as but special
cases of this more general class of supersymmetric extended object.  
These twelve points are the ones with $d\geq 2$ and denoted by $S$ in 
Table \ref{branescan}. For completeness, we have also included the 
superparticles with $d=1$ in $D=2,3,5$ and $9$. 

\begin{table}
$
\begin{array}{ccccccccccccccc}
~&D\uparrow&&&&&&&&&&&~\\
~&11&.&~&&S&&&T&&&&&~\\
~&10&.&V&S/V&V&V&V&S/V&V&V&V&V&~\\
~&9&.&S&&&&S&&&&&&~\\
~&8&.&~&&&S&&&&&&&~\\
~&7&.&~&&S&&&T&&&&&~\\
~&6&.&V&S/V&V&S/V&V&V&&&&&~\\
~&5&.&S&&S&&&&&&&&~\\
~&4&.&V&S/V&S/V&V&&&&&&&~\\
~&3&.&S/V&S/V&V&&&&&&&&~\\
~&2&.&S&&&&&&&&&&~\\
~&1&.&~&~&~&~&~&~&~&~&~&~&~\\
~&0&.&.&.&.&.&.&.&.&.&.&.&.~\\
~&~&0&1&2&3&4&5&6&7&8&9&10&11&d\rightarrow
\end{array}
$
\caption{ The brane scan, where $S$, $V$ and $T$ denote scalar, vector and
tensor multiplets.}
\la{branescan}
\end{table}

The letters $S$, $V$ and $T$ refer to {\it scalar}, {\it vector} and 
{\it tensor} 
respectively and describe the different kinds of particles that live on 
the worldvolume of the $p$-brane. Spin $0$ bosons and their spin 
$1/2$ fermionic partners are said to form a scalar supermultiplet. An 
example is provided by the eleven-dimensional supermembrane that 
occupies the $(D=11,d=3)$ slot on the branescan. It has $8$ spin $0$ 
and $8$ spin $1/2$ particles living on the three-dimensional (one 
time, two space) worldvolume  of the membrane. But 
as we shall see, it was subsequently realized that there also exist 
branes which have higher spin bosons on their worldvolume and belong 
to vector and tensor supermultiplets. 

A particularly interesting solution 
of eleven-dimensional supergravity, found 
by Bergshoeff, Sezgin and myself in collaboration 
with Chris Pope of Texas A\&M University, was called ``The membrane 
at the end of the universe.'' It described a four-dimensional spacetime
with the extra seven dimensions curled up into a seven-dimensional 
sphere and in which the supermembrane occupied the three-dimensional 
boundary of the four-dimensional spacetime (rather as the 
two-dimensional surface of a soap bubble encloses a three-dimensional 
volume).  This spacetime is of the kind first discussed earlier this 
century by the Dutch 
physicist Willem de Sitter and has a non-zero cosmological constant. 
It fact it is called anti-de Sitter space because the cosmological 
constant is negative. Now shortly after he first wrote down the 
equations of the membrane, Dirac pointed out in a (at the time 
unrelated) paper that anti-de Sitter space admits some strange kinds 
of fields that he called {\it singletons} which have no analogue in 
ordinary flat spacetime. These were much studied by 
Christian Fronsdal and collaborators at the University of California, 
Los Angeles, who pointed out that they reside not in the bulk of the 
anti-de Sitter space but on the three dimensional boundary. In 1988, 
the present author noted that, in the case of the seven-sphere 
compactification of eleven-dimensional supergravity, the singletons
correspond to the same $8$ spin 
$0$ plus $8$ spin $1/2$ scalar supermultiplet that lives on the 
worldvolume of the supermembrane, and it was natural to suggest that 
the membrane and the singletons should be identified. In this way, via 
the {\it the membrane at the end of the universe}, 
the physics in the bulk of the four-dimensional spacetime was really being 
determined by the physics on the three-dimensional boundary.  

Notwithstanding these and subsequent results, the
supermembrane enterprise ( Type $II$ A\&M Theory?) was ignored by most 
adherents of conventional superstring theory. Those who had worked on
eleven-dimensional supergravity and then on supermembranes spent the early
eighties arguing for {\it spacetime} dimensions greater than four, and the
late eighties and early nineties arguing for {\it worldvolume} dimensions
greater than two. The latter struggle was by far the more 
bitter\footnote{One string theorist I know would literally cover up 
his ears whenever the word ``membrane'' was mentioned within his 
earshot! Indeed, I used to chide my more conservative string theory 
colleagues by 
accusing them of being unable to utter the M-word. That the current 
theory ended up being called M-theory rather than Membrane theory was thus 
something of a Pyrrhic victory.}.

\section{Solitons, topology and duality}

Another curious twist in the history of supermembranes concerns their
interpretation as {\it solitons}. In their broadest definition, solitons
are classical solutions of a field theory corresponding to lumps of
field energy which are prevented from dissipating by a {\it topological}
conservation law, and hence display particle-like properties. The
classic example of a such soliton is provided by magnetic monopole
solutions of four-dimensional grand unified theories found by Gerard `t
Hooft of the the University of Utrecht in the Netherlands and Alexander
Polyakov, now at Princeton. Solitons play a ubiquitous role in
theoretical physics appearing in such diverse phenomena as condensed
matter physics and cosmology, where they are frequently known as {\it
topological defects}. 

To understand the meaning of a {\it topological conservation law}, we begin
by recalling that in 1917 the German mathematician Emmy Noether had shown
that to every global symmetry, there corresponds a quantity that is
conserved in time. For example, invariance under time translations, space
translations and rotations give rise to the laws of conservation of
energy, momentum and angular momentum, respectively. Similarly,
conservation of electric charge corresponds to a change in the phase of
the quantum mechanical wave functions that describe the elementary
particles. One might naively expect that conservation of magnetic charge
would admit a similar explanation but, in fact, it has a completely
different {\it topological} origin, and is but one example of what are now
termed topological conservation laws.  Topology is that branch of
mathematics which concerns itself just with the shape of things.
Topologically speaking, therefore, a teacup is equivalent to a doughnut
because each two-dimensional surface has just one hole: one can
continuosly deform one into the other. The surface of an orange, on the
other hand, is topologically distinct having no holes: you cannot turn an
orange into a doughnut. So it is with the intricate field configurations
describing  magnetic monopoles: you cannot turn a particle carrying $n$
units of magnetic charge into one with $n'$ units of magnetic charge, if
$n\neq n'$. Hence the charge is conserved but for topological reasons; not
for any reasons of symmetry.

In 1977, however, Claus Montonen of the University of Helsinki and David
Olive, now at the University of Wales at Swansea, made a bold conjecture.
Might there exist a {\it dual} formulation of fundamental physics in
which the roles of Noether charges and topological charges are reversed?
In such a dual picture, the magnetic monopoles would be the fundamental
objects and the quarks, W-bosons and Higgs particles would be the
solitons! They were inspired by the observation that in certain {\it
supersymmetric} grand unified theories, the masses $M$ of all the particles
whether elementary (carrying purely electric charge $Q$), solitonic
(carrying purely magnetic charge $P$) or {\it dyonic} (carrying both) are
described by a universal formula
\be
M^2=v^2(Q^2+P^2)
\ee
where $v$ is a constant. Note that the mass formula remains unchanged if
we exchange the roles of $P$ and $Q$! The Montonen-Olive conjecture was
that this electric/magnetic symmetry is a symmetry not merely of the mass
formula but is an exact symmetry of the entire quantum theory! The
reason why this idea remained merely a conjecture rather than a proof
has to do with the whole question of perturbative versus non-perturbative
effects. According to Dirac, the electric charge $Q$ is quantized in units
of $e$, the charge on the electron, whereas the magnetic charge
is quantized in units of $1/e$. In other words, $Q=me$ and $P=n/e$, where
$m$ and $n$ are integers. The symmetry suggested by Olive and Montonen
thus demanded that in the dual world, we not only exchange the integers
$m$ and $n$ but we also replace $e$ by $1/e$ and go from a regime of weak
coupling to a regime of strong coupling! This was very exciting because it
promised a whole new window on non-perturbative effects. On the other
hand, it also made a proof very difficult and the idea was largely
forgotten for the next few years.  

Although the original paper by Hughes, Liu and Polchinski made use of the
soliton idea, the subsequent impetus in supermembrane theory was to mimic
superstrings and treat the $p$-branes as fundamental objects in their own
right (analagous to particles carrying an electric Noether charge). 
Even within this framework, however, it was possible to
postulate a certain kind of duality between one $p$-brane and another by
relating them to the geometrical concept of {\it $p$-forms}.  (Indeed, this
is how $p$-branes originally got their name.) In their classic text
on general relativity, {\it Gravitation},  Misner, Thorne and Wheeler
\cite{Misner} provide a way to visualize {\it $p$-forms} as describing the
way in which surfaces are stacked. Open a cardbord carton containing a
dozen bottles, and observe the honeycomb structure of intersecting
north-south and east-west cardboard separators between the bottles.
That honeycomb structure of tubes is an example of a $2$-form in the
context of ordinary $3$-dimensional space. It yields a number (number
of tubes cut) for each choice of $2$-dimensional surface slicing
through the $3$-dimensional space. Thus a $2$-form is a device to
produce a number out of a surface. All of electromagnetism can be
summarized in the language of $2$-forms, honeycomb-like structures
filling all of $4$-dimensional spacetime. There are two such
structures, Faraday= $F$ and Maxwell= $*F$ each {\it dual}, or
perpendicular, to the other.  The amount of
electric charge or magnetic charge in an elementary volume is equal
respectively to the number of tubes of the Maxwell $2$-form $*F$ or
Faraday $2$-form $F$ that end in that volume. (In a world with no magnetic
monopoles, no tubes of $F$ would ever end.) To summarize, in
$4$-dimensional spacetime, an electric $0$-brane is dual to a magnetic
$0$-brane.

An equivalent way to understand why $0$-dimensional point particles produce
electric fields which are described by $2$-forms is to note that in
$4$-dimensional spacetime a pointlike electric charge can be surrounded by
a two-dimensional sphere.  Similarly, a string ($1$-brane) in
$4$-dimensional spacetime can be ``surrounded'' by a $1$-dimensional
circle, and so the electric charge per unit length of a string is
described by a $1$-form Maxwell field but its magnetic dual perpendicular
to it is described by a $3$-form Faraday field.  By contrast, in $5$
spacetime dimensions, although the Faraday field of a $0$-brane is still a
$2$-form, the dual Maxwell field is a now a $3$-form, consistent with the
fact that you now need a $3$-dimensional sphere to surround the pointlike
electric charge. But a $3$-form is just the Faraday field produced by a
string. Consequently, in $5$ spacetime dimensions, the magnetic dual of an
electric $0$-brane is a string.  Though in practice it is
harder to visualize, it is straightforward in principle to generalize this
duality idea to any $p$-brane in any spacetime dimension $D$. The
rule is that the Faraday field is a $(p+2)$ form and the dual Maxwell
field perpendicular to it is $(D-p-2)$-form.  Consequently,
the magnetic dual of an electric $p$-brane is a $\tilde
p$-brane where $\tilde p =D-p-4$. In particular, in the critical $D=10$
spacetime dimensions of superstring theory, a string ($p=1$) is dual to a
fivebrane ($\tilde p=5$). (If you have trouble imagining that in $10$
dimensions you need a $3$-dimensional sphere to surround a $5$-brane,
don't worry, you are not alone!)

Now the low energy limit of $10$-dimensional string theory is a
$10$-dimensional supergravity theory with a $3$-form Faraday field and dual
$7$-form Maxwell field, just as one would expect if the fundamental
object is a string. However, $10$-dimensional supergravity had one
puzzling feature that had long been an enigma from the point of view of
string theory. In addition to the above version there existed a {\it dual}
version in which the roles of the Faraday and Maxwell fields were
interchanged: the Faraday field was a $7$-form and the Maxwell field was
a $3$-form! This suggested to the present author in 1987, in analogy with
the Olive-Montonen conjecture, that perhaps this was indicative of a dual
version of string theory in which the fundamental objects are fivebranes! 
This became known as the {\it string/fivebrane duality conjecture}.  The
analogy was still a bit incomplete, however, because at that time the 
fivebrane was not regarded as a soliton.  

The next development came in 1988 when Paul Townsend of Cambridge
University revived the Hughes-Liu-Polchinski idea and showed that many of
the super $p$-branes also admit an interpretation as topological defects
(analogous to particles carrying a magnetic topological charge). Of
course, this involved generalizing the usual notion of a soliton: it need
not be restricted just to  a $0$-brane in four dimensions but might be
an extended object such as a $p$-brane in $D$-dimensions. Just like the
monopoles studied by Montonen and Olive, these solitons preserve half of
the spacetime supersymmetry and hence obey a relation which states that
their mass per unit $p$-volume is given by their topological charge.
Then in 1990, a major breakthrough for the string/fivebrane duality
conjecture came along when Strominger found that the
equations of the $10$-dimensional heterotic string admit a fivebrane as a
soliton solution which also preserves half the spacetime supersymmetry and
whose mass per unit $5$-volume is given by the topological charge
associated with the Faraday $3$-form of the string. Moreover, this mass
became larger, the smaller the strength of the string coupling, exactly as
one would expect for a soliton. He went on to suggest a complete
strong/weak coupling duality with the strongly coupled string
corresponding to the weakly coupled fivebrane.  By generalizing some
earlier work of Rafael Nepomechie (University of  Florida, Gainesville)
and Claudio Teitelboim (University of Santiago), moreover, it was
possible to to show that the electric charge of the fundamental string and
the magnetic charge of the solitonic fivebrane obeyed a Dirac quantization
rule. In this form, string/fivebrane duality was now much more closely
mimicking the electric/magnetic duality of Montonen and Olive. Then Curtis
Callan (Princeton University), Harvey and Strominger showed that similar
results also appear in both Type $IIA$ and Type $IIB$ string theories;
they also admit fivebrane solitons. However, since most physicists were
already sceptical of electric/magnetic duality in four dimensions, they
did not immediately embrace string/fivebrane duality in ten dimensions!  

Furthermore, there was one major problem with treating the fivebrane as
a fundamental object in its own right; a problem that has bedevilled
supermembrane theory right from the beginning: no-one knows how to
quantize fundamental $p$-branes with $p>2$.  All the techniques that
worked so well for fundamental strings and which allow us, for example, to
calculate how one string scatters off another, simply do not go through.
Problems arise both at the level of the worldvolume equations where our
old {\it bete noir} of non-renormalizability comes back to haunt us and
also at the level of the spacetime equations. Each term in string
perturbation theory corresponds to a two-dimensional worldsheet with more
and more holes: we must sum over all topologies of the worldsheet. But for
surfaces with more than two dimensions we do not know how to do this. 
Indeed, there are powerful theorems in pure mathematics which tell you
that it is not merely hard but impossible.  Of course, one could always
invoke the dictum that {\it God does not do perturbation theory}, but that
does not cut much ice unless you can say what He does do!  So there were
two major impediments to string/fivebrane duality in $10$ dimensions.
First, the electric/magnetic duality analogy was ineffective so long as
most physicists were sceptical of this duality. Secondly, treating
fivebranes as fundamental raised all the unresolved issues of 
non-perturbative quantization.  

The first of these impediments was removed, however, when Ashoke Sen (Tata
Institute) revitalized the Olive-Montonen conjecture by establishing that
certain dyonic states, which their conjecture demanded, were indeed
present in the theory. Many duality sceptics were thus converted. 
Indeed this inspired Nathan Seiberg (Rutger's University) and Witten to
look for duality in more realistic (though still supersymmetric)
approximations to the standard model. The subsequent industry, known as
Seiberg-Witten theory, provided a wealth of new information on
non-perturbative effects in four-dimensional quantum field theories,
such as quark-confinement and symmetry-breaking, which would have been
unthinkable just a few years ago.

The Olive-Montonen conjecture was originally intended to apply to
four-dimensional grand unified field theories. In 1990, however,  Anamarie
Font, Luis Ibanez, Dieter Lust and Fernando Quevedo at CERN and,
independently, Soo Yong Rey (University of Seoul) generalized the idea to
four-dimensional superstrings, where in fact the idea becomes even more
natural and goes by the name of $S$-duality. 

In fact, superstring theorists had already become used to a totally
different kind of duality called $T$-duality. Unlike,
$S$-duality which was a non-perturbative symmetry and hence still
speculative, $T$-duality was a perturbative symmetry and rigorously
established. If we compactify a string theory on a circle then,
in addition to the Kaluza-Klein particles we would expect in an ordinary
field theory, there are also extra {\it winding}
particles that arise because a string can wind around the circle.
$T$-duality states that nothing changes if we exchange the roles of the
Kaluza-Klein and winding particles provided we also exchange the radius
of the circle $R$ by its inverse $1/R$. In short, a string cannnot tell
the difference between a big circle and a small one!

\section{String/string duality in six dimensions}

Recall that when wrapped around a circle, an $11$-dimensional membrane
behaves as if it were a $10$-dimensional string. In a series of papers
between 1991 and 1995, a team at Texas A\&M University involving  
Ramzi Khuri, James T. Liu, Jianxin Lu, Ruben Minasian, Joachim Rahmfeld,
and myself argued that this may also be the way out of the problems of
$10$-dimensional string/fivebrane duality. If we allow four of the ten
dimensions to be curled up and allow the solitonic fivebrane to wrap
around them, it will behave as if it were a $6$-dimensional 
solitonic string! The fundamental string will remain a fundamental string
but now also in $6$-dimensions. So the $10$-dimensional string/fivebrane
duality conjecture gets replaced by a $6$-dimensional string/string
duality conjecture. The obvious advantage is that, in contrast to the
fivebrane, we do know how to quantize the string and hence we can put the
predictions of string/string duality to the test. For example, one can
show that the coupling constant of the solitonic string is indeed
given by the inverse of the fundamental string's coupling constant, in
complete agreement with the conjecture. 

When we spoke of string/string duality, we originally had in mind a
duality between one heterotic string and another, but the next major
development in the subject came in 1994 when Christopher Hull (Queen Mary
and Westfield College, London University) and Townsend suggested that, if
the four-dimensional compact space is chosen suitably, a six-dimensional
heterotic string can also be dual to a six-dimensional Type $IIA$ string!    
These authors also added futher support to the idea that the Type 
$IIA$ string originates in eleven dimensions.

It occurred to the present author that string/string duality has another
unexpected pay-off. If we compactify the six-dimensional spacetime on two
circles down to four dimensions, the fundamental string and the
solitonic string will each acquire a $T$-duality. But here is the miracle:
the $T$-duality of the solitonic string is just the $S$-duality of the
fundamental string, and vice-versa! This phenomenon, in which the
non-perturbative replacement of $e$ by $1/e$ in one picture is just the
perturbative replacement of $R$ by $1/R$ in the dual picture, goes by the
name of {\it Duality of Dualities}.  See Table 3.  Thus four-dimensional
electric/magnetic duality, which was previously only a conjecture, now
emerges automatically if we make the more primitive conjecture of
six-dimensional string/string duality.

\begin{table}
$
\begin{array}{lll}
&{\bf Fundamental \, string}&{\bf Dual \, string}\\
&&\\
{\bf T-duality}: & Radius \leftrightarrow 1/(Radius) & charge \leftrightarrow
1/(charge)\\ &Kaluza-Klein \leftrightarrow Winding & Electric
\leftrightarrow Magnetic\\ {\bf S-duality}: & charge 
\leftrightarrow 1/(charge) &
Radius \leftrightarrow 1/(Radius) \\ & Electric \leftrightarrow Magnetic &
Kaluza-Klein \leftrightarrow Winding \end{array}
$
\la{Duality}
\caption{Duality of dualities}
\end{table}

\section{M-theory}

All this previous work on $T$-duality, $S$-duality, and string/string
duality was suddenly pulled together under the umbrella of $M$-theory by
Witten in his, by now famous, talk at the University of Southern
California in February 1995. Curiously enough, however, Witten still
played down the importance of supermembranes. But it was only a matter of
time before he too succumbed to the conclusion that we weren't doing just
string theory any more! In the coming months, literally hundreds of
papers appeared in the internet confirming that, whatever $M$-theory may
be, it certainly involves supermembranes in an important way.  For
example, in 1992 R. G\"{u}ven (Bosporus University) had shown that
eleven-dimensional supergravity admits a solitonic fivebrane solution dual
to the fundamental membrane solution found the year before by Stelle and
myself.  See the $(D=11,d=6)$ point marked by a $T$ in Table 
\ref{branescan}. It did not take long to realize that
$6$-dimensional string/string duality (and hence $4$-dimensional
electric/magnetic duality) follows from $11$-dimensional
membrane/fivebrane duality. The fundamental string is obtained by wrapping
the membrane around a one-dimensional space and then compactifying on a
four-dimensional space; whereas the solitonic string is obtained by
wrapping the fivebrane around the four-dimensional space and then
compactifying on the one-dimensional space. Nor did it take long before
the more realistic kinds of electric/magnetic duality envisioned by
Seiberg and Witten were also given an explanation in terms of
string/string duality and hence $M$-theory.

Even the chiral $E_8 \times E_8$ string, which according to Witten's
earlier theorem could never come from eleven-dimensions, was given an
eleven-dimensional explanation by Petr Horava (Princeton University) and
Witten. The no-go theorem is evaded by compactifying not on a circle (which
has no ends), but on a line-segment (which has two ends).  It is ironic 
that having driven the nail into the coffin of eleven-dimensions (and
having driven Gell-Mann to utter ``Ugh!''), Witten was the one to pull
the nail out again! He went on to argue that if the size of this
one-dimensional space is large compared to the six-dimensional Calabi-Yau
manifold, then our world is approximately five-dimensional. This may have
important consequences for confronting $M$-theory with experiment. For
example, it is known that the strengths of the four forces change with
energy. In supersymmetric extensions of the standard model, one finds that
the fine structure constants $\alpha_3,\alpha_2,\alpha_1$ associated with
the $SU(3) \times SU(2) \times U(1)$  all meet at about $10^{16}$ GeV,
entirely consistent with the idea of grand unification.  The strength of
the dimensionless number $\alpha_G=GE^2$, where $G$ is Newton's contant
and $E$ is the energy, also almost meets the other three, but not quite.
This near miss has been a source of great interest, but
also frustration.  However, in a universe of the kind envisioned by
Witten, spacetime is approximately a narrow five dimensional layer bounded
by four-dimensional walls. The particles of the standard model live on the
walls but gravity lives in the five-dimensional bulk. As a result, it is
possible to  choose the size of this fifth dimension so that all four
forces meet at this common scale. Note that this is much
less than the Planck scale of $10^{19}$ GeV, so gravitational effects may
be much closer in energy than we previously thought; a result that would
have all kinds of cosmological consequences.

Thus this eleven-dimensional framework now provides the starting point for
understanding a wealth of new non-perturbative phenomena, including
string/string duality, Seiberg-Witten theory, quark confinement,
particle physics phenomenology and cosmology.

\section{Black holes and $D$-branes}

Type $II$ string theories differ from heterotic theories in one important
respect: in addition to the usual Faraday $3$-form charge, called the
Neveu-Schwarz charge after Andre Neveu (University of Montpelier) and
Schwarz, they also carry so-called Ramond charges, named after Pierre
Ramond of the University of Florida, Gainesville. These are associated with
Faraday $2$-forms and $4$-forms in the case of Type $IIA$ and Faraday
$3$-forms and $5$-forms in the case of Type $IIB$. Accordingly in 1993,
Jiaxin Lu and I were able to find new solutions of the Type $IIA$
string equations describing super $p$-branes with $p=0,2$ and their
duals with $p=6,4$ and new solutions of Type $IIB$ string equations with
$p=1,3$ and their duals with $p=5,3$.  Interestingly enough, the Type
$IIB$ superthreebrane is {\it self-dual}, carrying a magnetic charge
equal to its electric charge.  This meant that there were more points
on the brane-scan than had previously been appreciated. These occupy 
the $V$ slots in Table \ref{branescan}.
For all these solutions, the mass per unit $p$-volume was given by the
charge, as a consequence of the preservation half of the spacetime
supersymmetry.  However, we recognized that they were in fact just the
extremal mass=charge limit of more general non-supersymmetric solutions
found previously by Horowitz and Strominger. These solutions, whose mass
was greater than their charge, exhibit {\it event horizons}: surfaces from
which nothing, not even light, can escape. They were {\it black} branes!

Thus another by-product of these membrane breakthroughs has been an
appreciation of the role played by black holes in particle physics and
string theory. In fact they can be regarded as black branes wrapped
around the compactified dimensions. These black holes are tiny $(10^{-35}$
meters) objects; not the multi-million solar mass objects that are
gobbling up galaxies. However, the same physics applies to both and there
are strong hints by Lenny Susskind (Stanford University) and others that 
M-theory may
even clear up many of the apparent paradoxes of quantum black holes raised
by Hawking.

As we have already discussed, one of the biggest unsolved mysteries in
string theory is why there seem to be billions of different ways of
compactifying the string from ten dimensions to four and hence
billions of competing predictions of the real world. Remarkably,
Brian Greene of Cornell University, David Morrison of Duke University and
Strominger have shown that these wrappped around black branes actually
connect one Calabi-Yau vacuum to another. This holds promise of a
dynamical mechanism that would explain why the world is as it is, in other
words, why we live in one particular vacuum. A fuller discussion may 
be found in Greene's book \cite{Greene}.

Another interconnection was recently uncovered by Polchinski who realized 
that the Type $II$ super $p$-branes carrying Ramond charges may be
identified with the so-called Dirichlet-branes (or $D$-branes, for
short) that he had studied some years ago by looking at strings with
unusual boundary conditions. Dirichlet was a French mathematician who first
introduced such boundary conditions. These $D$-branes are just the
surfaces on which open strings can end. In the process, he discovered an
$8$-brane in Type $IIA$ theory and a $7$-brane and $9$-brane in Type $IIB$
which had previously been overlooked. See Table \ref{branescan}. This 
$D$-brane
technology has opened up a whole new chapter in the history of
supermembranes.  In particular, it has enabled Strominger and Cumrun Vafa
from Harvard to make a comparison of the black hole entropy calculated
from the degeneracy of wrapped-around black brane states with the
Bekenstein-Hawking entropy of an extreme black hole. Their agreement
provided the first microscopic explanation of black hole entropy.
Moreover, as Townsend had shown earlier, the extreme black hole solutions
of the ten-dimensional Type $IIA$ string (in other words, the Dirichlet
$0$-branes) were just the Kaluza-Klein particles associated with wrapping
the eleven-dimensional membrane around a circle. Moreover,
four-dimensional black holes also admit the interpretation of intersecting
membranes and fivebranes in eleven-dimensions.  All this holds promise of
a deeper understanding of black hole physics via supermembranes.

\section{Eleven to twelve: is it still too early?}

We have remarked that eleven spacetime dimensions are the maximum allowed
by super $p$-branes.  This is certainly true if we believe that the
Universe has only one time dimension. Worlds with more than one time
present all kind of headaches for theoretical physicists and they
prefer not to think about them.  For example, there would be no 
``before" and ``after" in the conventional sense.  Just for fun,
however, in 1987 Miles Blencowe (Imperial College, University of London)
and I imagined what would happen if one relaxed this one-time requirement.
We found that we could not rule out the possibility of a supersymmetric
extended object with a ($2$ space, $2$ time) worldvolume living in a
($10$ space, $2$ time) spacetime. We even suggested that the Type $IIB$
string with its ($1$ space, $1$ time) worldsheet living in a
($9$ space, $1$ time) spacetime might be descended from this object in much
the same way that the Type $IIA$ string with its ($1$ space, $1$ time)
worldsheet living in a ($9$ space, $1$ time) spacetime is descended from
the  ($2$ space, $1$ time) worldvolume of the supermembrane living in a
($10$ space, $1$ time) spacetime.  This idea lay dormant for almost a
decade but has recently been revived by Vafa and others in the context of
{\it $F$-theory}.  The utility of $F$-theory is certainly beyond dispute:
it has yielded a wealth of new information on string/string duality. But
should the twelve dimensions of $F$-theory be taken seriously? And if so,
should $F$-theory be regarded as more fundamental than $M$-theory? (If $M$
stands for Mother, maybe $F$ stands for Father.) To make sense of
$F$-theory, however, it seems necessary to somehow freeze out the twelfth
timelike dimension where there appears to be no dynamics. Moreover,
Einstein's requirement that the laws of physics be invariant under changes
in the spacetime coordinates seems to apply only to ten or eleven of the
dimensions and not to twelve. So the symmetry of the theory, as far as we
can tell, is only that of ten or eleven dimensions.  The more conservative
interpretation of $F$-theory, therefore, is that the twelfth dimension is
just a mathematical artifact with no profound significance.  Time (or
perhaps I should say ``Both times'') will tell.

\section{So what is M-theory?}

Is M-theory to be regarded literally as membrane theory? In other words 
should we attempt to ``quantize'' the eleven dimensional membrane in 
some, as yet unknown, non-perturbative way? Personally, I think the jury is 
still out on whether this is the right thing to do. Witten, for example, 
strongly
believes that this is not the correct approach. He would say, in physicist's 
jargon, that we
do not even know what the right degrees of freedom are. So although
$M$-theory admits $2$-branes and $5$-branes, it is probably much more
besides. 

Recently, Tom Banks and Stephen Shenker at Rutgers together with Willy
Fischler from the Univerity of Texas and Susskind 
have even proposed a rigorous definiton of $M$-theory known as
M(atrix) theory which is based on an infinite number of Dirichlet
$0$-branes. In this picture spacetime is a fuzzy concept in which the
spacetime coordinates $x,y,z,...$ are matrices that do not commute e.g.
$xy \neq yx.$ This approch has generated great excitement but does yet
seem to be the last word. It works well in high dimensions but 
as we descend in dimension it seems to break down before we reach the real
four-dimensional world.

Another interesting development has recently been provided by Juan 
Maldacena at Harvard, who has suggested that $M$-theory on anti-de 
Sitter space, including all 
its gravitational interactions, may be completely described by a 
non-gravitational 
theory on the boundary of anti-de Sitter space. This holds promise not 
only of a deeper understanding of M-theory, but may also throw light on 
non-perturbative aspects of the theories that live on the boundary, 
which in some circumstances can include the kinds of quark theories 
that govern the strong nuclear interactions. Models of this kind, 
where a bulk theory with gravity is equivalent to a boundary theory 
without gravity, 
have also been advocated by `t Hooft and independently by Susskind 
who call them {\it holographic} theories. The reader may notice a 
striking similarity to the earlier idea of ``The membrane at the end 
of the universe'' \cite{Duffsutton} and interconnections between the 
two are currently being explored.

$M$-theory has sometimes been called the {\it Second 
Superstring Revolution}, but we feel this 
is really a misnomer. It certainly involves new ideas every bit as 
significant as those of the 1984 string revolution, but its reliance 
upon supermembranes and eleven dimensions makes it is sufficiently 
different from traditional string theory to warrant its own name. One cannot 
deny the tremendous historical influence of the last decade of 
superstrings on our current perspectives. Indeed, it is the pillar 
upon which our belief in a quantum consistent M-theory rests. In my opinion, 
however, the 
focus on the perturbative aspects of one-dimensional objects
moving in a ten-dimensional spacetime that prevailed during this period will 
ultimately  be seen to be a small (and perhaps physically insignificant) 
corner
of $M$-theory. The overriding problem in superunification in the coming years
will be to take the Mystery out of $M$-theory, while keeping the Magic and
the Membranes. 

\section{Acknowledgements}

I am grateful for correspondence with Freeman Dyson, Robert Low, Ergin 
Sezgin and Edward Witten.



\end{document}